\documentclass[%
reprint,
superscriptaddress,
 amsmath,amssymb,
 aps,
prb,
]{revtex4-2}

\usepackage{graphicx}
\usepackage{dcolumn}
\usepackage{bm}
\usepackage{xcolor}
\usepackage[caption=false]{subfig}
\usepackage{float}

\begin{document}


\title{On the role played by electrons in the stress-strain curves of ideal crystalline solids}

\author{Margherita Marsili}%
\affiliation{Department of Physics and Astronomy, University of Bologna, Viale Carlo Berti Pichat 6/2, Bologna, 40127, Italy}
\email{margherita.marsili@unibo.it; clelia.righi@unibo.it}

\author{Elisa Damiani}%
\affiliation{Department of Physics and Astronomy, University of Bologna, Viale Carlo Berti Pichat 6/2, Bologna, 40127, Italy}

\author{Davide Dalle Ave}%
\affiliation{Department of Physics and Astronomy, University of Bologna, Viale Carlo Berti Pichat 6/2, Bologna, 40127, Italy}

\author{Gabriele Losi}%
\affiliation{Department of Physics and Astronomy, University of Bologna, Viale Carlo Berti Pichat 6/2, Bologna, 40127, Italy}

\author{M. Clelia Righi}
\affiliation{Department of Physics and Astronomy, University of Bologna, Viale Carlo Berti Pichat 6/2, Bologna, 40127, Italy}
\email{clelia.righi@unibo.it}

\date{\today}

\begin{abstract}
The mechanical properties of a solid, which relate its deformation to external applied forces, are key factors in enabling or disabling the use of an otherwise optimal material in any application, strongly influencing also its service lifetime. Intrinsic crystal deformation mechanisms, investigated experimentally on single crystals with low dislocation densities, have been studied theoretically through atomistic simulations, mainly focusing on lattice-induced instabilities.
Here, instead, we employ density functional theory and a thermodynamic analysis to probe and analyze 
the way in which the electronic charge of crystalline solids (Cu, Al and diamond) responds to uniaxial strain and affects their mechanical properties. Indeed, despite the very simple nature of our models, and in the presence of minimal atomic displacements, we find that the stress strain curves of Cu and Al deviate from a simple linear elastic behavior. Within a thermodynamics perspective, the features of such curves can be interpreted in terms of first and second order phase transitions, which originate from Van-Hove singularities of the electronic density of states crossing the Fermi level and electron redistribution within the solid, respectively.

\end{abstract}

\maketitle

\section{Introduction}

\begin{figure*}[hbt!]
\includegraphics[width=\textwidth]{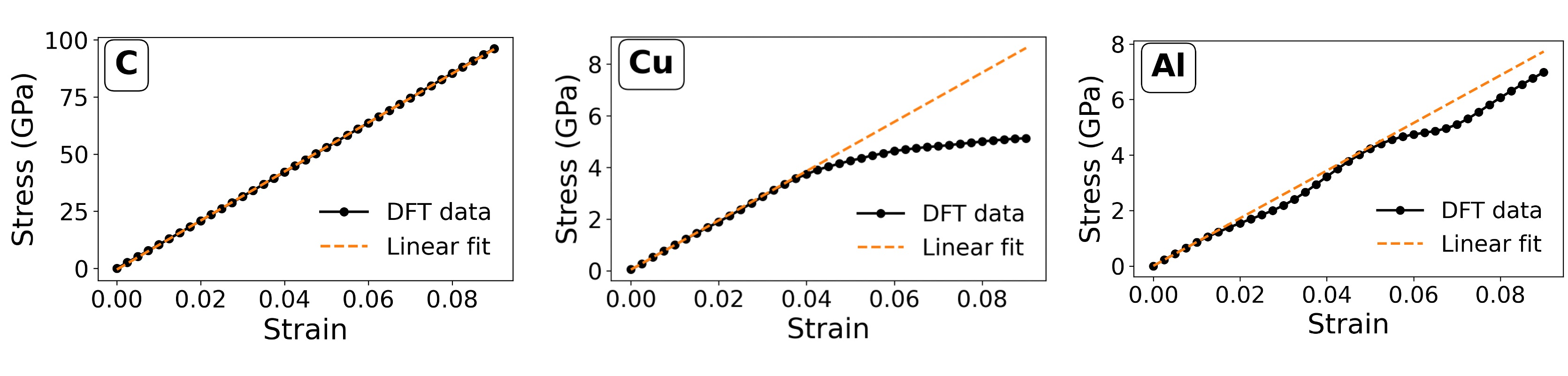}
\caption{\label{fig:stress-strain}Stress-strain curve for diamond (left), copper (center), aluminum (right). The linear regions are fitted with
linear regression. Young’s moduli are calculated as $E = \sigma_{zz}/\epsilon_z$ in the linear region. $\sigma_{zz}$ is the stress associated to the uniaxial strain along the z-direction. (See S.I.)}
\end{figure*}

The mechanical response of a solid, its capability of resisting or accommodating applied strains or stresses through reversible or irreversible deformations, dictates the range of mechanical stimuli the solid is able to comply with, directly influencing all its possible applications.
Whenever loads, external forces, are applied to solid objects, they will deform. In the elastic regime the object will return to its original shape and size after load removal, whereas the plastic regime is characterized by irreversible deformations.
This macroscopic, mechanical behavior of materials is clearly visible in the so called stress-strain curves, widely used reference graphs in material science and manufacturing, where the stress within a material is plotted against the relative change in length (or volume), namely the strain, it is opposing to.\\
In a generic ideal stress-strain plot under compressive load, it is possible to identify the elastic region, where the curve follows a linear trend, and, beyond it, the yield point that traditionally marks the transition  between the linear-elastic regime and the plastic one where irreversible atomic planes sliding begins \cite{callister_book}.
While at the macroscopic and mesoscopic levels the fingerprint of this behavior is displayed by cracks, dislocations, deformation bands, folds, shear markings and finally phase transitions, \cite{he_scirep_2022,bespalova_russ_phys_j_2017}, at the microscopic, atomistic, level it may be anticipated by different kinds of processes ultimately leading to instabilities \cite{Lion_PhysRevMaterials2022}: phonon modes softening \cite{Grimvall_RevModPhys2012,Dubois_PhysRevB2006, Clatterbuck_PhysRevLett2003,Roundy_phil_mag_A}, bonds formation or destabilization \cite{Jiang_nature_2013,li_pnas2012}. Indeed pressure has been shown to alter the chemical properties of elements, even literally subverting the periodic table \cite{Dong_PNAS2022,Hou_JACS2025}.\\

Although most materials have a polycristalline character for which it is the defect structure (grain boundaries, dislocations, slip planes) that forces the way in which they will deform, single crystals with low dislocation densities, for which elastic/plastic deformations are due to purely intrinsic crystal deformations, are the best playground to explore basic deformation mechanisms. And, besides experiments, it is in this limiting case of perfect crystals that microscopic atomistic simulations can play a role providing further mechanistic insights on the mechanical behaviour of materials. But while most of the studies focus on the properties and on the response of the atomic lattice itself as the primary source of instability and deformation path \cite{Dubois_PhysRevB2006,Clatterbuck_PhysRevLett2003,Roundy_phil_mag_A,Cogollo-Olivo_PhysRevB2018,marini_PhysRevB2012}, the role and what directly happens to the electronic charge distribution has seldomly been addressed.\\

Here, we employ density functional theory (DFT) to directly probe the response of the electronic degrees of freedom of crystalline solids to uniaxial strain and look at how it affects their mechanical response. To do this we focus on the behaviour of crystalline Al, Cu and diamond bulks. To be able to identify basic electronic mechanisms, disentangled as much as possible from the ionic degrees of freedom, we make use of minimal models made up of few atoms per unit cell and not presenting defects, impurities, or flaws.
Despite the extremely simplified nature of our systems, the simulated stress-strain curves of the metals display the typical trend of the macroscopic stress-strain curves that feature elastic to plastic regime transition. 
Moreover, our result show that the changes in the mechanical response of the solid, associated to the deviation from the linear behaviour of the stress-strain curve, correspond to minimal deviations form the ideal geometry. Looking at the stress strain curve in a thermodynamics perspective, such critical points are signatures of crossovers, eventually leading to phase transition in the thermodynamic limit. We thus try to interpret such crossovers in terms of changes in the electronic structure. Our analysis show, on one side, that the critical points of the stress-strain curve, associated to second order crossovers, are linked to changes in the amount of charge that the increasing load "moves" within and between the atomic planes. 
On the other side, in the case of Al, the features of its mechanical response between 6\% and 9\% strain, associated to a first order phase transition crossover, can be interpreted as a Lifshitz transition driven by density of states (DOS) Van-Hove singularities crossing the Fermi level.\\
The paper is organized as follows: after that computational details are presented, the stress-strain behaviour of the three solids will be analysed; then such stress-strain curves will be interpreted in a thermodynamics perspective and critical points linked to first and second order cross-overs will be identified; finally the physical origin of such criticalities will be explored linking the critical points present in the stress-strain curves to changes in electronic charge distribution within and between atomic planes, and to Van-Hove singularities crossing the Fermi level.  

\section{Computational details}
DFT calculations have been carried out as implemented in the Quantum Espresso (QE) package \cite{Giannozzi_jphys_cond_mat_2009,Giannozzi_j_phys_cond_mat_2017} using ultrasoft pseudopotential \cite{Vanderbilt-1990} and plane waves expansion.
For the exchange-correlation functional, the Perdew-Burke-Ernzerhof (PBE) \cite{PBE-1996} was chosen, adopting the parameterization by Rabe-Rappe-Kaxiras-Joannopoulos \cite{rappe_PhysRevB1990}. 18$\times$18$\times$18, 12$\times$12$\times$12, and 16$\times$16$\times$16 Monkhorst-Pack grids were used to sample the Brillouin zone for Al, Cu, and diamond, respectively. The kinetic energy cut-off of 40 Ryd (90 Ryd), for the wavefunctions, and 320 Ryd (720 Ry), for the charge density were employed for aluminum and diamond (copper). 
Forces were minimized employing variable cell calculations, optimizing at the same time both atomic positions and cell edges. 
The three bulk materials are simulated as cubic cells with the $xy$ plane parallel to the crystallographic (001) direction. 
For Al DOS calculations a 121$\times$121$\times$121 k-point grid was used, employing the tetrahedron method to better describe Van-Hove singularities.     
To simulate the effect of load, we strained the cells of the three materials along the z-direction performing an accurate planar relaxation (xy cartesian plane) of the cell structures while allowing a full relaxation in all directions of the  atomic positions. 
The remaining uniaxial stress along the vertical axis is derived from the DFT final stress tensor. 

\section{Stress-strain behaviour}

\begin{figure*}[htb!]
 \centering
  \subfloat[]{\includegraphics[width=0.33\textwidth]{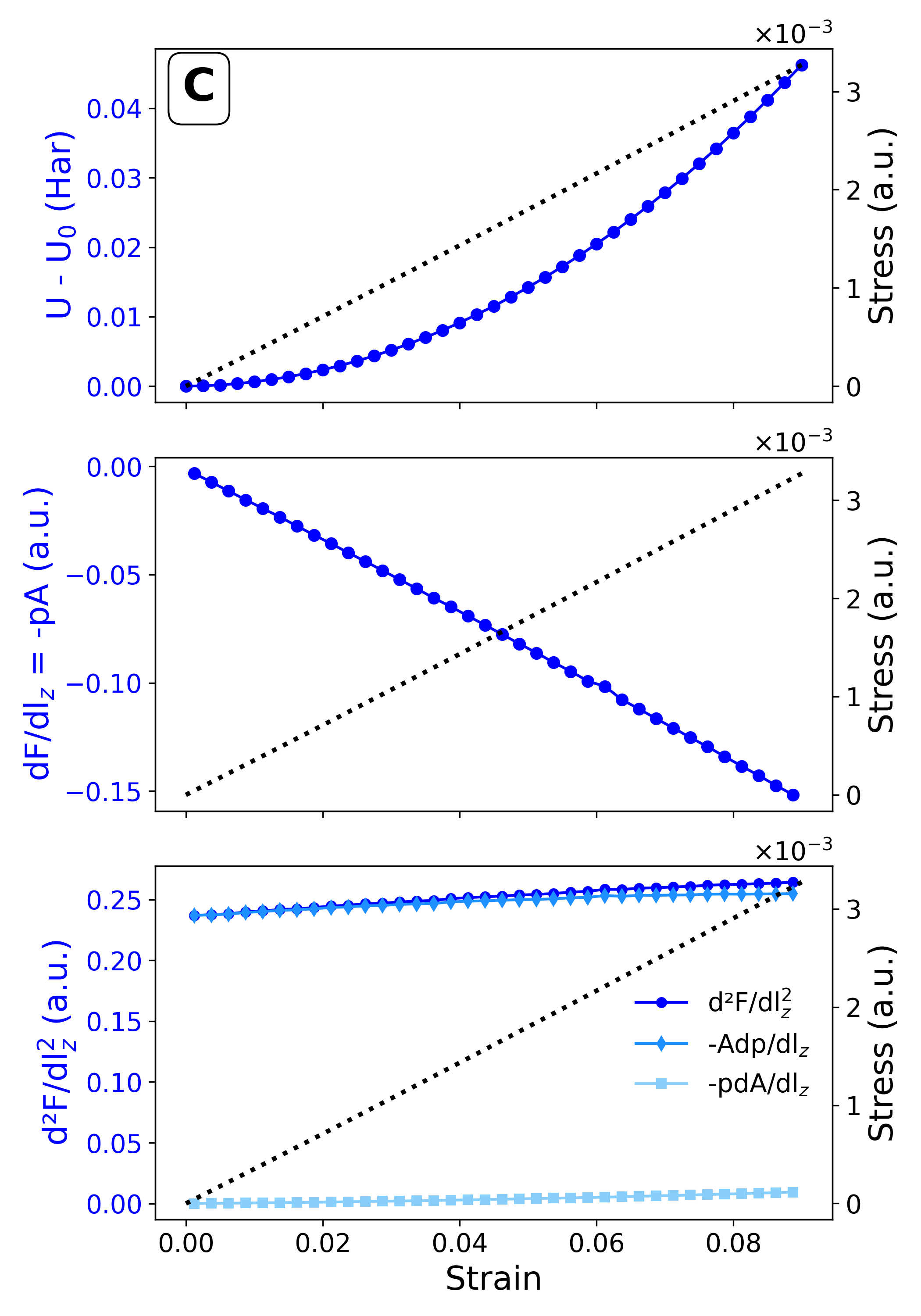}}
\subfloat[]{\includegraphics[width=0.33\textwidth]{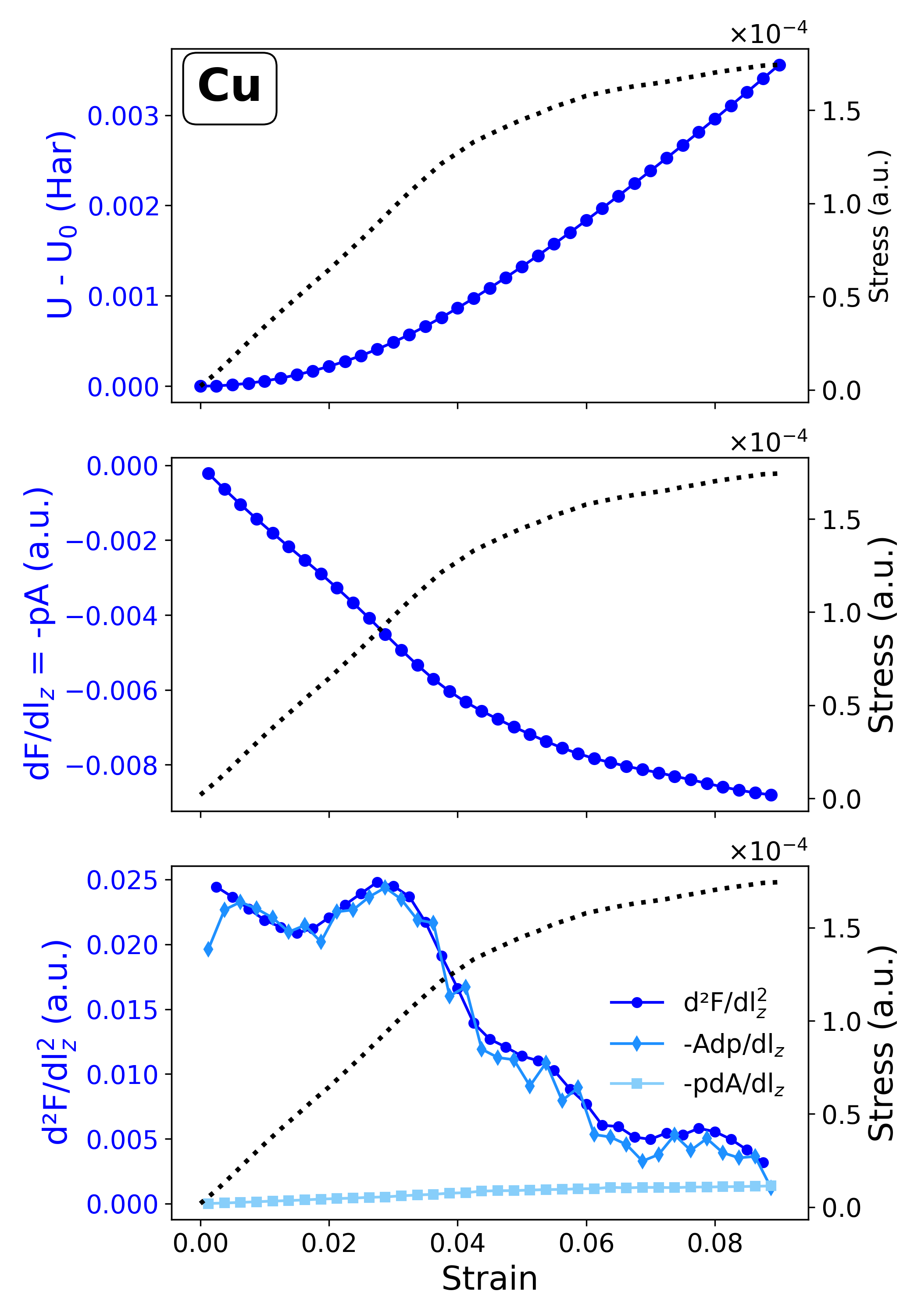}}
\subfloat[]{\includegraphics[width=0.33\textwidth]{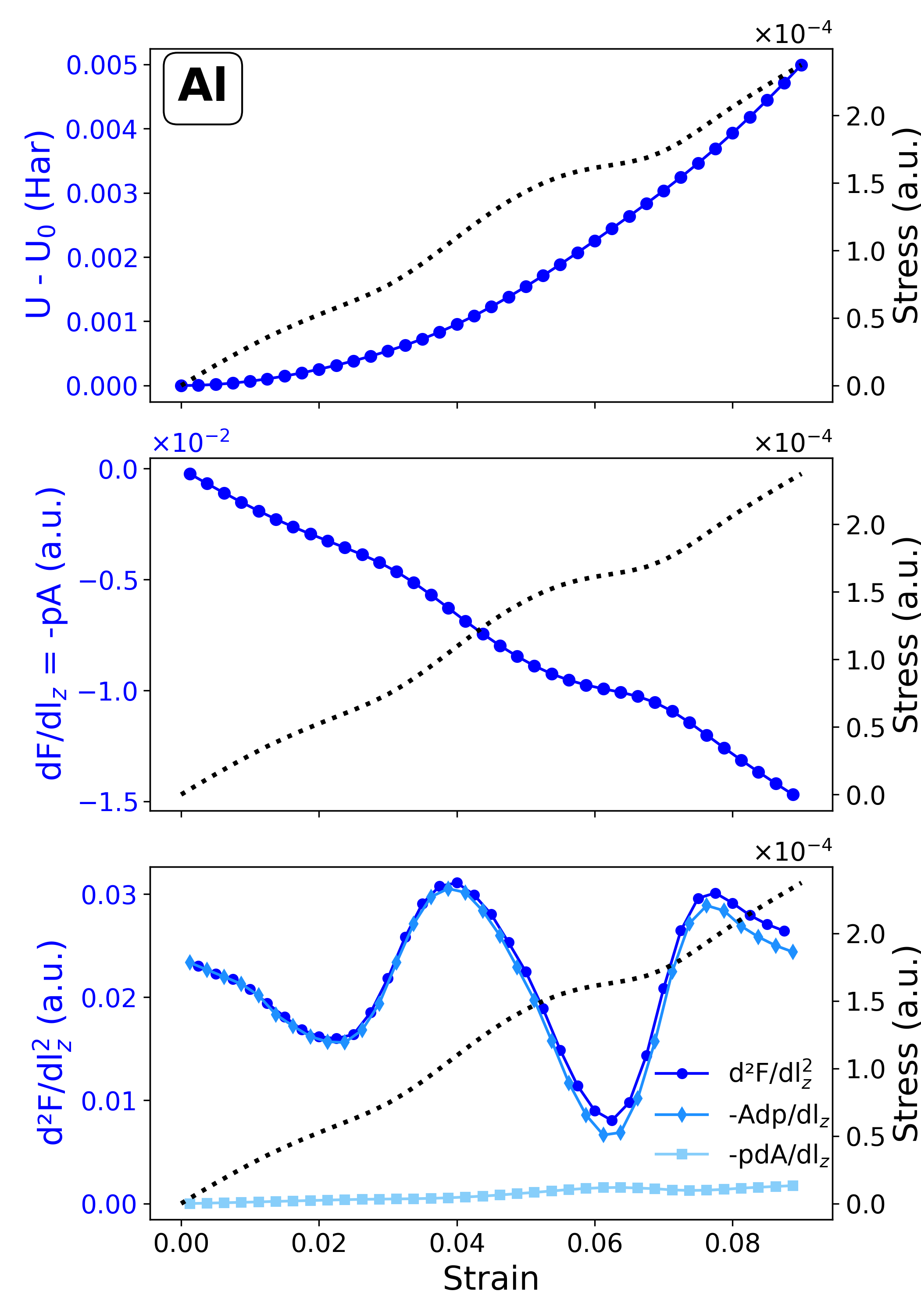}}%
\caption{(a) diamond; (b) Cu; (c) Al. From top to bottom: Helmholtz free energy $F$; its first order derivative $\left(\frac{\partial F}{\partial l_z}\right)$; $\left(\frac{\partial^2 F}{\partial l_z^2}\right)_T=T_1+T_2$, $T_1=-A\left(\frac{\partial P}{\partial l_z}\right)_T$ and $T_2=-p\left(\frac{\partial A}{\partial l_z}\right)_T$. In all panel stress values are reported as a black dotted line and refer to the right axis labels.\label{fig:termo_potentials} }
\end{figure*}

Pressure is applied to the solids by reducing the bulk cells dimension along the z-direction at the same time allowing the cell to deform within the xy plane, and all the atoms to fully relax.
The cell dimension along $z$, $l$, was varied, by applying vertical strains $\epsilon_z=\frac{l-l_0}{l_0}$ ranging from -0.01 to -0.09 ($l_0$ being the equilibrium value). The details of the structural modifications of the three solids, such as area, volume changes, and deviations from ideal atomic positions are reported in the S.I.. Importantly, despite the fact that full relaxation is allowed, concerning the atomic positions no significant displacement is found at any strain even for the two metals. 

After constraining the unit cell dimension along the z direction to progressively compressed values and having allowed the full optimization of the cell in the xy plane,  $\sigma_{zz}$ is the only non-null element left in the stress tensor of the crystal. By collecting $\sigma_{zz}$ for increasing strains $\epsilon_z$, the stress-strain curves for the three materials, shown in Fig. \ref{fig:stress-strain}, are built.

The stress-strain curves of the three materials differ qualitatively: while, within diamond, stress grows linearly in the whole strain range, in the case of Cu it grows fairly linearly up to 3\%-4\% strain, slightly lowering its slope between 0.01 and 0.02 strain, however at 0.03 strain it starts bending in a fashion that resembles a macroscopic elastic-plastic transition, where, beyond the yield point, the same increase of strain generates a lower increase of stress in the material: a sort of material softening. The case of Al presents an even more complex behavior: stress grows linearly up to $\sim$3\% strain, from that point and up to $\sim$5.5\%  strain, stress is still increasing linearly but with a higher slope, as if the material were hardening. The curve then becomes nearly flat up to $\sim$7\%, when it starts again to grow linearly.

The diamond stress-strain curve very well compares with that computed for uniaxial strain along the (001) direction on a 1$\times$1 cell in Ref. \cite{luo_JPCC2010}.
Interestingly in experiments Aluminum presents an extremely early onset of non linear elastic behaviour  and the yield point is difficult to identify. Indeed, experimental stress-strain curves of Al single crystals \cite{romanova_metals2022,becker_met_transA1991,kahn_intjplasticity2015,bell_phil_mag1967,Ha_math_mech_sol2011,basson_acta_materialia2000} show a bending behaviour with decreasing slope already in the region close to 0 strain. Importantly in some of the curves \cite{romanova_metals2022,kahn_intjplasticity2015}, between $\sim$2\% and $\sim$4\% strain, a sudden increase of the slope of the stress-strain curve is also found, whereas others \cite{becker_met_transA1991,basson_acta_materialia2000} show an hardening of the crystal 
following a region where the response is flat.
From the knowledge of the stress-strain curves and of the structural deformation of the three materials, their elastic properties, namely their Young’s modulus and Poisson ratio, can be obtained. In all cases they compare well with experiments as shown in Tab.\ref{tab:elastic_properties} (further details can be found in the S.I.).   

\begin{table}[ht!]
    \centering
    \begin{tabular}{c|cc|cc}
    \hline \hline
         Material &  $E$ (GPa) &  $E_{exp}$ (GPa) &  $\nu$ &  $\nu_{exp}$\\
         \hline
         C  & 1068  & 700-1200 & 0.107  & 0.10-0.30\\
         Cu & 97    & 110 & 0.341  & 0.34\\
         Al & 82    & 69 & 0.326  & 0.33\\
         \hline \hline
    \end{tabular}
    \caption{Comparison between calculated and experimental values of Young’s modulus $E$ and Poisson’s ratio $\nu$ for Al, Cu, C. Experimental values are obtained from Ref.\cite{callister_book}.}
    \label{tab:elastic_properties}
\end{table}

\section{Critical points thermodynamic analysis}
\begin{figure*}[htb!]
\includegraphics[width=\textwidth]{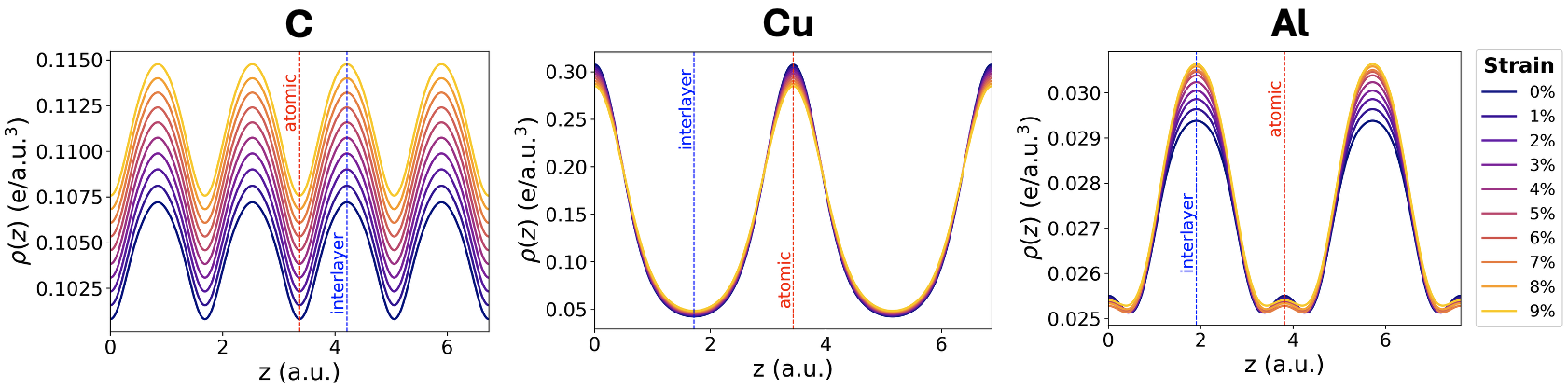}%
\caption{\label{fig:eq_planar_average} C (left), Cu (center), and Al (right) charge density $\rho$ planar average profiles along the [001] direction for selected strain values. The position of an atomic plane is shown by a red-dashed line, whereas a middle position between two atomic planes is shown by a blue-dashed line.} 
\end{figure*}
To gain insight and differentiate the critical points present in the stress-strain curve it is possible to take a thermodynamic perspective: in our simulations, at fixed 0K temperature, the bulk cell is strained on the $z$-direction (by fixing the cell parameter $l_z$), allowing the basis area $A$ to vary in order to minimize the total energy. As a result the system is left with an internal pressure $P$ that originates from the internal forces that would restore the equilibrium geometry. The thermodynamic potential of choice is then the Helmholtz free energy, $F(T,l_z)=U-TS$, where the thermodynamic variable conjugated to $l_z$ is the total force acting along the $z$-direction $f_z=PA$ that keeps the strained system at equilibrium \cite{blundell_blundell_book}.
The differential of this thermodynamic potential would then be
$dF=-SdT-f_z\cdot l_z=-SdT-pAdl_z$, its first order derivative at fixed temperature $\left(\frac{\partial F}{\partial l_z}\right)_T=-pA$, and its second order derivative
\begin{equation}
 \left(\frac{\partial^2 F}{\partial l_z^2}\right)_T=-A\left(\frac{\partial P}{\partial l_z}\right)_T-p\left(\frac{\partial A}{\partial l_z}\right)_T=T_1+T_2.  
\end{equation}
$T_1=-A\left(\frac{\partial P}{\partial l_z}\right)_T$ represents the contribution to $\left(\frac{\partial^2 F}{\partial l_z^2}\right)_T$ of the variation of pressure with respect to $l_z$, whereas $T_2=-p\left(\frac{\partial A}{\partial l_z}\right)_T$ is the contribution coming from the variation of the basis area.

The Helmholtz free energy, and its first and second order derivatives with respect to strain are shown for diamond, together with the stress-strain curve, in Fig. \ref{fig:termo_potentials}(a), all three quantities are smoothly varying with strain, reflecting the unperturbed linear behaviour of the entire stress-strain curve. 
In Fig.\ref{fig:termo_potentials}(b), the same quantities are shown for Cu. The Helmholtz free energy, which at $T=0$ is nothing but the total energy of the system, shown in the top panel of Fig. \ref{fig:termo_potentials}(b), is continuous; its first derivative, second panel, resembles closely the stress-strain curve, implying that the basis area $A$ is a smooth function of $l_z$. The second order derivative of the thermodynamic potential, shown in the bottom panel of Fig.  \ref{fig:termo_potentials}(b), presents a discontinuity, corresponding to the point where the stress-strain curve (and $\left(\frac{\partial F}{\partial l_z}\right)_T$) bends. Looking at the light blue dots of the same figure, it is clear how the discontinuity in $\left(\frac{\partial^2 F}{\partial l_z^2}\right)_T$ arises from the term $T_1$, i.e. from an abrupt change in the slope of the pressure, in agreement with the behavior of the stress-strain curve.

In the case of Al, shown in Fig. \ref{fig:termo_potentials}(c), the situation is more complex, as anticipated by the behavior of the stress-strain curve. Like in the case of Cu, the first derivative of the thermodynamic potential $\left(\frac{\partial F}{\partial l_z}\right)_T$  (displayed in the central panel) closely resembles the stress-strain curve and, when further differentiating the thermodynamic potential, the main contribution arises from the variation of pressure with respect to $l_z$. However in this case we are in the presence of an oscillating behavior: around $\epsilon_z \sim 0.03$, where the stress-strain curve changes its slope, the $\left(\frac{\partial^2 F}{\partial l_z^2}\right)_T$ has a sudden jump, a discontinuity similar to the one present for Cu, but in the opposite direction, i.e. increasing the value of $\left(\frac{\partial^2 F}{\partial l_z^2}\right)_T$; subsequently, by applying further strain, $\left(\frac{\partial^2 F}{\partial l_z^2}\right)_T$ immediately and quickly starts to decrease reaching an absolute minimum at $\epsilon_z \sim 0.0625$, within the region where the stress-strain curve becomes flat.
The way in which the critical point locations are identified is detailed in the S.I..  

The thermodynamics perspective allows a better characterization and classification of the critical points of the stress-strain curves: the critical point of Cu manifests itself
as a discontinuity in the second order derivative of the thermodynamic potential, it can thus be classified as a crossover linked to a second order phase transition happening at the bending points of the stress-strain diagram (and of the Poisson's ratio one, see S.I.). 
In the case of Al we are instead in the presence of two different kind of critical points/regions.
Around  3\% strain, the discontinuity of $\left(\frac{\partial^2 F}{\partial l_z^2}\right)_T$ can be associated to a crossover, eventually leading to second order phase transition in the thermodynamic limit, consistently to what is happening for Cu, but in the opposite direction. 
The region around 6\% strain instead looks more similar to a crossover leading, in the thermodynamic limit, to a first order phase transition: the $pA$ plot (central panel of Fig. \ref{fig:termo_potentials}(c)) shows that, while varying the thermodynamic variable $l_z$ its conjugate variable $pA$ is nearly constant. A similar behaviour is found in text books concerning the isothermal compression of a gas below its critical temperature, in this case, during the crossover, as the volume is reduced, pressure stays constant until all the vapour has condensed.

\section{Electronic charge redistribution}
\begin{figure}[htb!]
\includegraphics[width=0.45\textwidth]{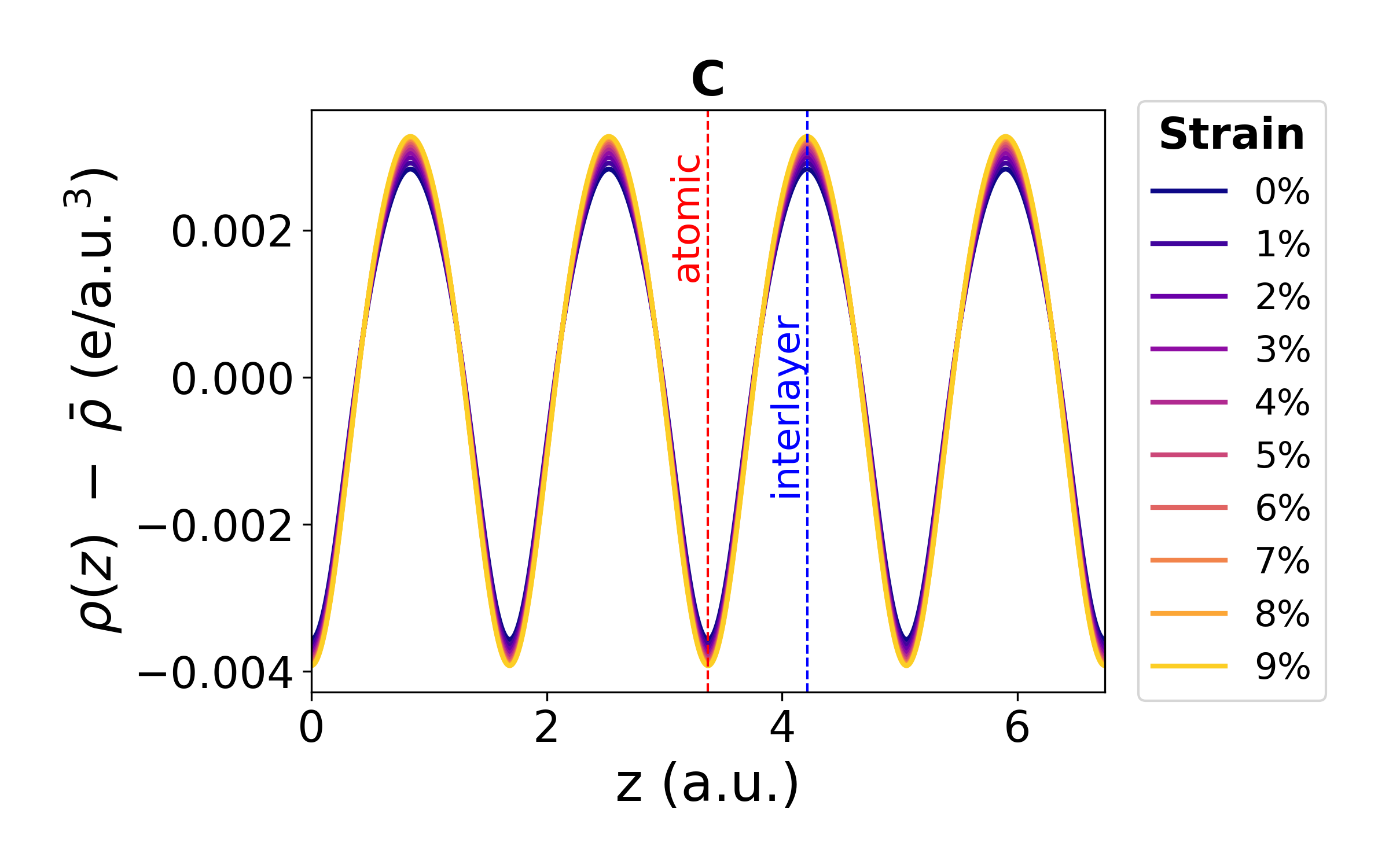}%
\caption{\label{fig:planar_average_relative} Diamond charge density $\rho$ planar average profiles along the [001] direction shifted by $\bar{\rho}=N_{el}/V$.  An atomic plane and an interlayer plane are marked by red and blue dashed lines respectively. $N_{el}$ is the number of electrons of the supercell, while $V$ is the supercell volume.} 
\end{figure}
The minimal nature of our models naturally quenches the lattice distortion degree of freedom, moreover the basis area is a smooth and a well behaved function of $\epsilon_z$ (see S.I.).
Therefore the physical origin of the critical points in the stress strain curve, sign of first or second order phase transitions, must be sought within the electronic system. We thus analyze the way the electronic charge redistributes itself upon increasing load.\\
Planar averages of charge density $\rho(z)$ are depicted in Fig.\ref{fig:eq_planar_average} as line profiles along the [001] direction. The position of atomic planes and interlayer planes are shown for all the slabs by red and blue dashed lines respectively.
By comparing the charge densities profiles of the three materials at equilibrium (corresponding to the $\epsilon_z=0$ lines in Fig.\ref{fig:eq_planar_average}), a main differences between aluminum and diamond on one side, and copper on the other is manifest:
whereas charge peaks are found between atomic planes
for aluminum and diamond, in the case of copper charge maxima are located at the atomic planes.
When load is applied, charge density at the atomic planes of Al does not vary significantly whereas the excess charge, due to compression, is accumulated between atomic planes. On the other hand in Cu, there is a depletion of charge at the atomic planes and correspondingly a slight accumulation of charge between planes. Diamond apparently has a completely different behaviour: in its case the overall charge profile is rising. However, by removing the contribution of the average density $\bar{\rho}=N_{el}/V$, $V$ being the volume of the supercell, it is clear that, just like Al, diamond is slightly accumulating charge between atomic planes as shown in  Fig.\ref{fig:planar_average_relative}. 

To spot possible peculiar behavior of the electronic charge at the critical points, modifications of the charge profiles are monitored at each strain increment, i.e. upon going from $\epsilon_z^{i-1}$ to $\epsilon_z^i$, being $\epsilon_z^0=0$ and $\epsilon_z^{i_{MAX}}=0.09$.  Namely we compute the charge profile increments:
$\Delta \rho(z,(\epsilon_z^i+\epsilon_z^{(i-1)})/2)=\rho(z,\epsilon_z^i)-\rho(z,\epsilon_z^{(i-1)})$. 
\begin{figure}[htb!]
\includegraphics[width=\linewidth]{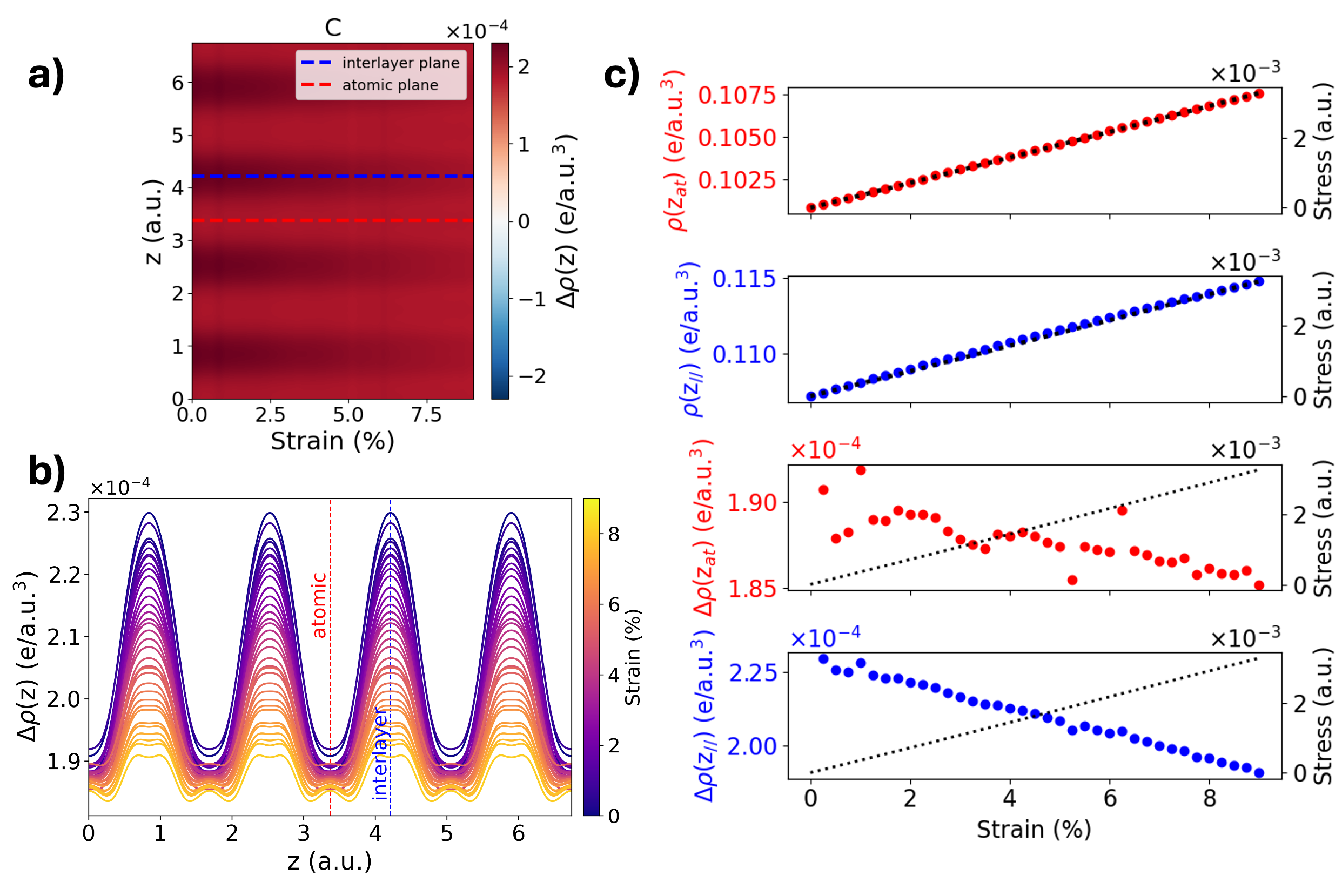}%
\caption{\label{fig:charge_analysis_C} (a) Charge profile increments $\Delta \rho(z,\epsilon_z)$ as a function of strain $\epsilon_z$ in the horizontal axis, and $z$ in the vertical axis. The positions of an interlayer and an atomic plane are marked by a dashed blue and red line, respectively. (b) Charge profile increments $\Delta \rho(z,\epsilon_z)$ as a function of $z$ for all the computed strain values. (c) From top to bottom: planar average charge densities profiles computed at atomic planes $\rho(z_{at}, \epsilon)$; values of the planar average charge densities profile computed at interlayer planes $\rho(z_{\parallel}, \epsilon)$; charge profile increments computed at atomic planes, i.e. cut of the heatmap (a) along the red-dashed line; charge profile increments computed at interlayer planes, i.e. cut of the heatmap (a) along the blue-dashed line. All panels refer to diamond.}
\end{figure}

Charge profiles increments, $\Delta \rho(z,\epsilon)$, are shown Fig.\ref{fig:charge_analysis_C}(b) for the case of diamond. 
In this case, both at the atomic planes and interlayer planes charge is increasing almost linearly with strain, as shown in the top two panels of the Fig.\ref{fig:charge_analysis_C}(c). The corresponding local charge increments, bottom two panels, are always positive and slightly constantly decreasing, as also attested by the "fading" heatmap (Fig.\ref{fig:charge_analysis_C}(a)). 

\begin{figure}[htb!]
\includegraphics[width=\linewidth]{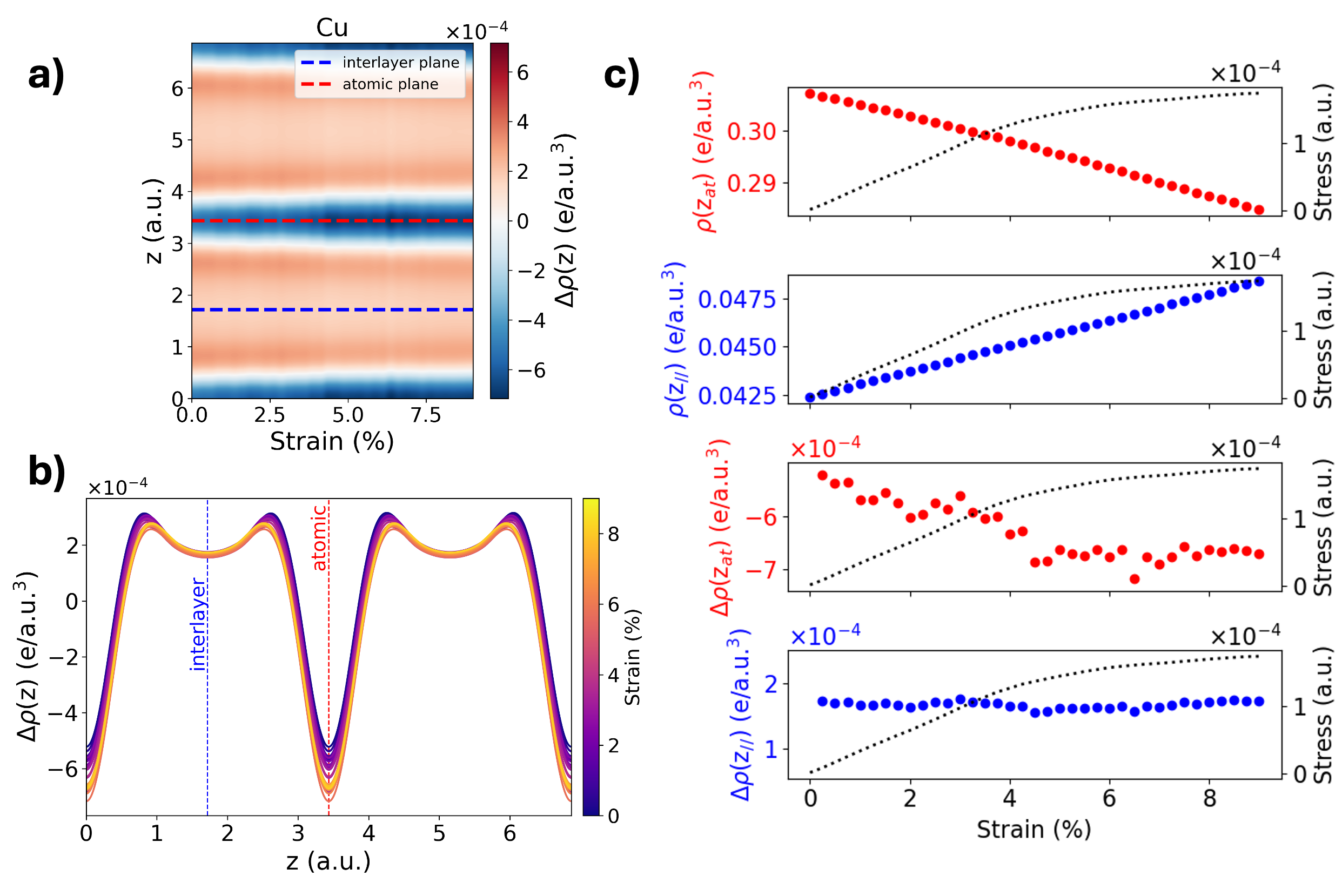}%
\caption{\label{fig:charge_analysis_Cu} (a) Charge profile increments $\Delta \rho(z,\epsilon_z)$ as a function of strain $\epsilon_z$ in the horizontal axis, and $z$ in the vertical axis. The positions of an interlayer and an atomic plane are marked by a dashed blue and red line, respectively. (b) Charge profile increments $\Delta \rho(z,\epsilon_z)$ as a function of $z$ for all the computed strain values. (c) From top to bottom: planar average charge densities profiles computed at atomic planes $\rho(z_{at}, \epsilon)$; values of the planar average charge densities profile computed at interlayer planes $\rho(z_{\parallel}, \epsilon)$; charge profile increments computed at atomic planes, i.e. cut of the heatmap (a) along the red-dashed line; charge profile increments computed at interlayer planes, i.e. cut of the heatmap (a) along the blue-dashed line. All panels refer to Cu.}
\end{figure}

The results for Cu  are collected in Fig.\ref{fig:charge_analysis_Cu}. At the atomic planes charge is decreasing piece wise linearly with
strain (top panel of Fig.\ref{fig:charge_analysis_Cu}(c)). At the same time, at the interlayer planes, the electronic charge is linearly increasing (second top panel of Fig.\ref{fig:charge_analysis_Cu}(c)). Consistently, charge increments within the interlayer planes (bottom panel) are nearly constant, whereas, within the atomic planes, (second bottom plot), $\Delta \rho(z_{at},(\epsilon_z))$ abruptly drops from $\sim -5.5\cdot 10^{-4} e/\AA{}^3$
to $\sim -7\cdot 10^{-4} e/\AA{}^3$ between 3\% and 4\% strain, signaling an increase of charge depletion rate (with strain) in that region. Remarkably, the drop in $\Delta \rho(z_{at},(\epsilon_z))$ is
accompanied by the softening in the material, as shown by the stress-strain curve plotted for reference in red in the same figure. The discontinuous change of the rate of charge depletion at the atomic plane is thus marking the second order phase transition identified at $\epsilon_z=3\%$ in the thermodynamics analysis.\\ 
\begin{figure}[htb!]
\includegraphics[width=\linewidth]{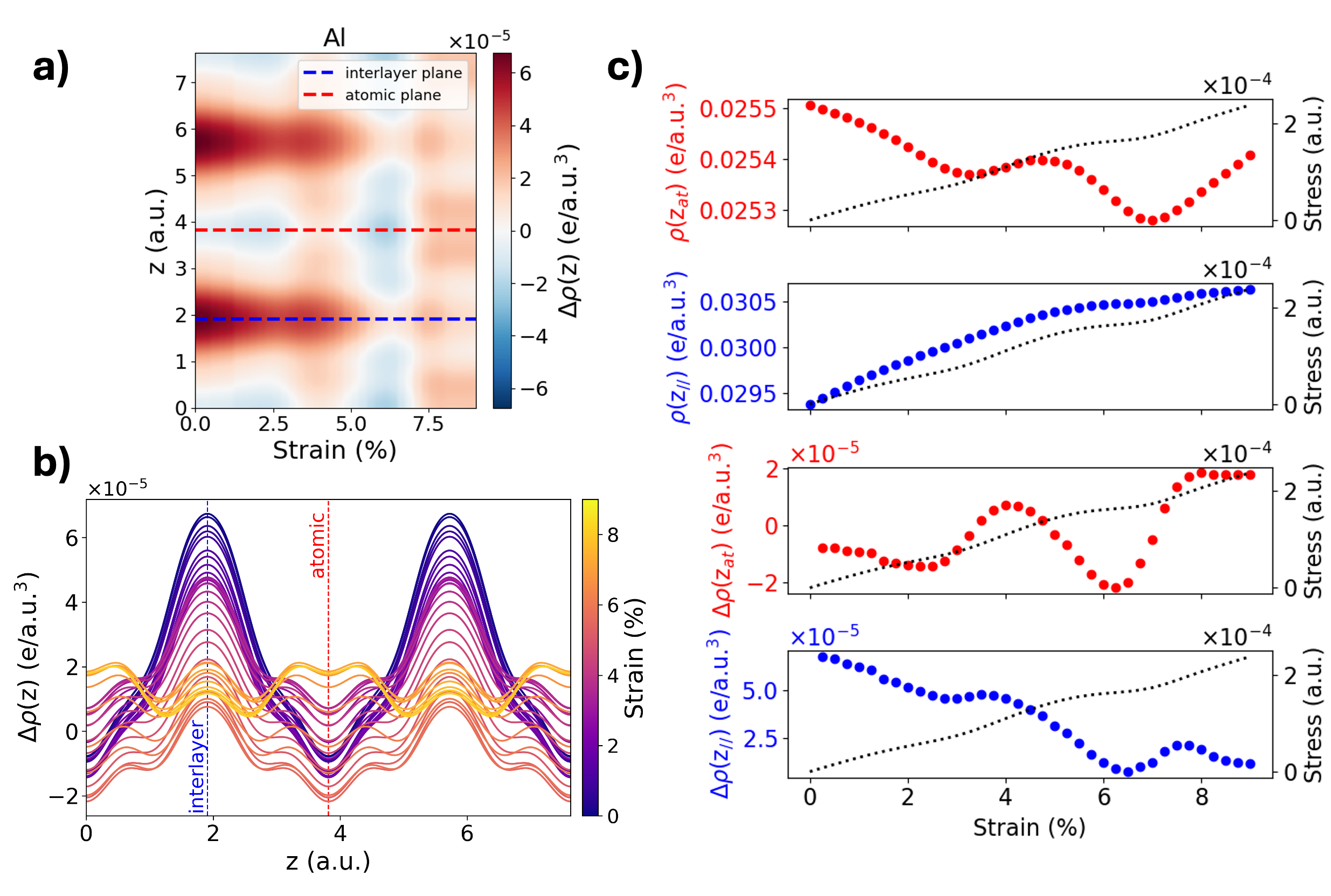}%
\caption{\label{fig:charge_analysis_Al} (a) Charge profile increments $\Delta \rho(z,\epsilon_z)$ as a function of strain $\epsilon_z$ in the horizontal axis, and $z$ in the vertical axis. The positions of an interlayer and an atomic plane are marked by a dashed blue and red line, respectively. (b) Charge profile increments $\Delta \rho(z,\epsilon_z)$ as a function of $z$ for all the computed strain values. (c) From top to bottom: planar average charge densities profiles computed at atomic planes $\rho(z_{at}, \epsilon)$; values of the planar average charge densities profile computed at interlayer planes $\rho(z_{\parallel}, \epsilon)$; charge profile increments computed at atomic planes, i.e. cut of the heatmap (a) along the red-dashed line; charge profile increments computed at interlayer planes, i.e. cut of the heatmap (a) along the blue-dashed line. All panels refer to Al.} 
\end{figure}

The complexity of Al stress-strain curve is reflected also in the evolution of its charge profiles and of the charge increments, collected in Fig.\ref{fig:charge_analysis_Al}. Within the interlayer planes (Fig.\ref{fig:charge_analysis_Al}(c), second panel from the top), charge consistently grows in the entire strain range with changes in slope that reflect those of the stress-strain curve. In contrast, at the atomic planes the, overall decreasing, charge features oscillations that clearly follow those of the second derivative of the thermodynamic potential (bottom panel of Fig.\ref{fig:termo_potentials}(c)). 

Looking at the charge increments at the two critical points in the two bottom panels of Fig.\ref{fig:charge_analysis_Al}(c), we see that the second order phase transition crossover at $\epsilon_z=0.03$ is characterized by a discontinuity of the charge increments at the atomic plane and nearly constant increments at the interlayer planes, resembling the situation of Cu. It is interesting to note that the sign of the discontinuity correlates with the mechanical behaviour: in the Cu case, there is a negative jump of the increments and the material is softening, whereas in Al there is a positive jump and the material hardens.  
The first order phase transition crossover at $\epsilon_z\sim 6\%$ is instead marked by a deep minimum of both the local charge increments.

\section{Al DOS evolution with strain}
\begin{figure}
    \centering
    \includegraphics[width=1\linewidth]{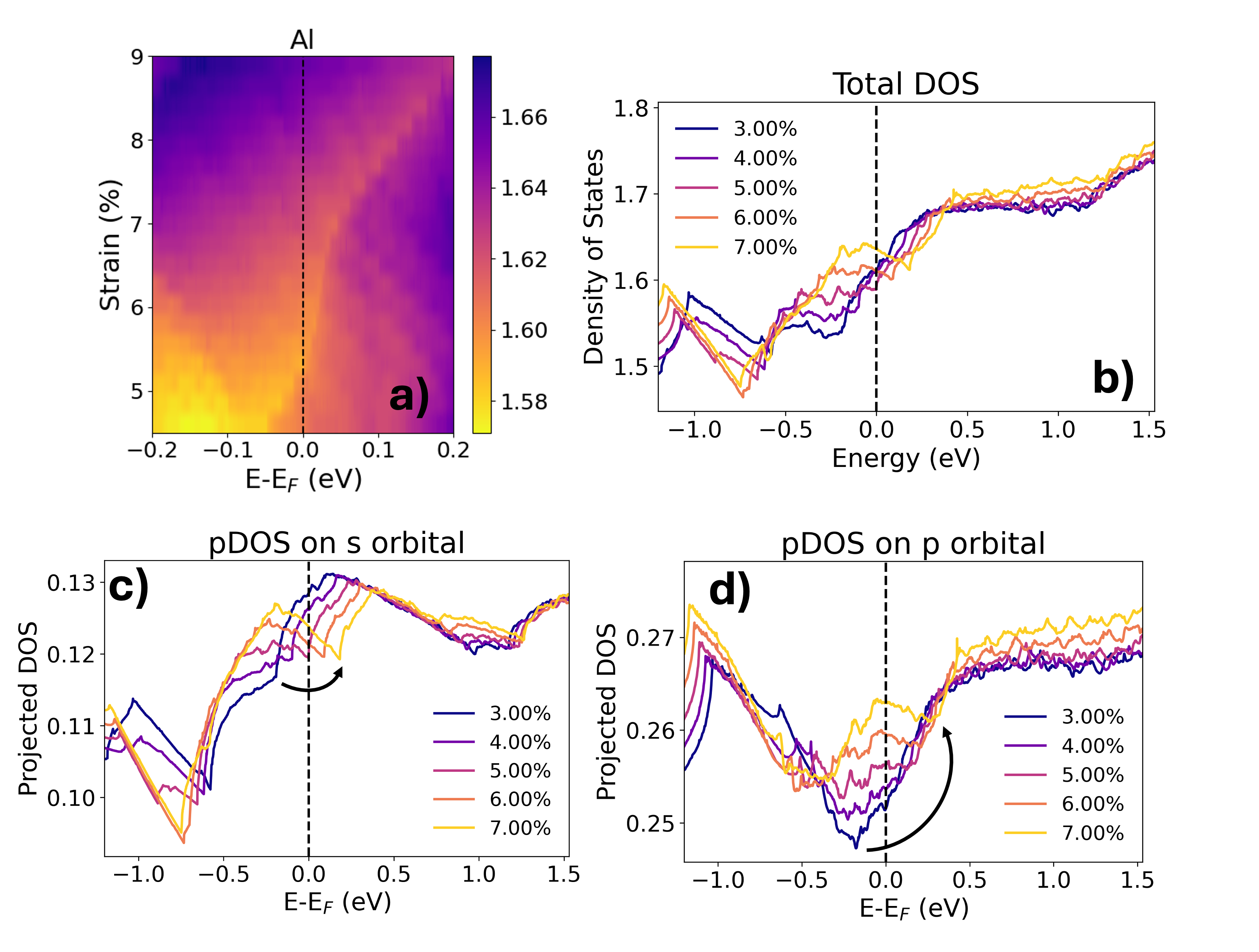}
    \caption{(a) Al DOS close to the Fermi level for strain values beteween 4.5\% and 9.0\%. (b) Al total DOS for selected strain value around the critical value of $\epsilon_z=0.06$. (c) Al DOS projected on the s-orbitals for selected strain value around the critical value of $\epsilon_z=0.06$. (d) Al DOS projected on the p-orbitals for selected strain value around the critical value of $\epsilon_z=0.06$. }
    \label{fig:dos_al_0.06_strain}
\end{figure}

It is common knowledge that, within solidification from a liquid phase to a solid phase (a first order phase transition) volume changes do not produce any pressure change because the system is rearranging its internal structure. 
Similarly, in the case of the Al crystal around $\epsilon_z \sim 6\%$, the increase of strain is not producing any change in $pA$, the force by which the solid is responding. In analogy with gas condensation, this points towards the presence of an internal rearrangement of its degrees of freedom, provided in this case by its electronic structure. 
We have thus analyzed the evolution of Al electronic DOS with increasing uniaxial load, to see whether the crossover at $\sim6\%$ strain could be driven by the sudden availability or unavailability of novel electronic states.  Indeed, pressure is able to modify the band structures of materials, and when new bands begin crossing the Fermi level, or bands previously crossing it are entirely pushed below or above it, the Fermi surface changes its topology, a phenomenon known as Lifshitz transition \cite{lifshitz1960}. Lifshitz transition are typically marked by Van-Hove singularities crossing the Fermi level and, as a consequence, by discontinuities of the density of states at the Fermi Level. Importantly, in the case of Al under hydrostatic pressure, a Lifhsitz transition was detected by nuclear magnetic resonance \cite{Meissner_2014}.\\
Fig.\ref{fig:dos_al_0.06_strain}(a) shows the evolution of Al DOS for energies close to the Fermi level, marked by a dashed black line, for strain values between 4.5\% and 9.0\%. A minimum of the DOS can be clearly seen crossing the Fermi level between 5.5\% and 6.5\% strain. The DOS profiles for selected strains, Fig.\ref{fig:dos_al_0.06_strain}(b) show this feature, which is due to both a contribution due to $s$ electrons (Fig.\ref{fig:dos_al_0.06_strain}(c)) and $p$ electrons (Fig.\ref{fig:dos_al_0.06_strain}(d)). Importantly, as shown in the S.I., around the critical point $\epsilon_z \sim 3\%$, the Al DOS close to the Fermi level does not display any significant change.\\ 
It can be thus concluded that the first order crossover happening, for Al, around $\epsilon_z \sim 6\%$ is linked to Van-Hove singularities crossing the Fermi level and altering the availability of electronic states for the response of the solid to pressure.    

\section{Conclusions}
In conclusion we employed DFT to study the response of the electronic system of crystalline Al, Cu and diamond to uniaxial strain.
In order to quench lattice deformation degrees of freedom and focus on basic electronic mechanisms, we make use of minimal models made up of few atoms per unit cell and not presenting defects, impurities, or flaws. Young's moduli and Poisson ratio, computed in the limit of low strains, agree well with experimental values.
Surprisingly, despite the very simple nature of our models, and in the presence of minimal atomic displacements, the stress strain curves of Cu and Al deviate from a simple linear elastic behavior. 

Within a thermodynamic perspective two distinct features can be distinguished in the stress-strain curve of the two metals: 
in one case (at $\sim$4\% strain for Cu and $\sim$3\% strain for Al) the change in the slope of the stress-strain curve is a manifestation of the discontinuity of the second order derivative of the thermodynamic potential. In the second case, 
happening in the region around $\sim$6\% strain for Al, the stress-strain curve is almost flat, at same time the first order derivative of the thermodynamic potential, the thermodynamic variable $pA$, conjugated to $l_z$ , is substantially constant as $l_z$ varies, resembling crossover leading, in the thermodynamic limit, to a first order phase transition. Besides providing a better classification and characterization of the critical points of the stress strain curve, the thermodynamics analysis confirms the fundamental role played by the charge depletion/accumulation at the atomic and interlayer planes. Indeed simple bending of the stress-strain curve, associated to the softening (for Cu) and hardening (for Al) of the materials, and linked to sudden decrease for Cu and increase 
for Al of $\left(\frac{\partial^2 F}{\partial l_z^2}\right)_T$ - the rate of change with strain of $pA$, the force by which the unit cell of the crystal opposes to the compression - is accompanied by a sudden increase in atomic plane charge depletion
for Cu and, viceversa, a sudden increase in atomic plane charge accumulation for Al. 
In the meanwhile, charge accumulation at the interlayer planes stays nearly constant. \\
The second kind of structure, where the stress-strain curve of Al is nearly flat, is a 
sort of extreme softening of the material; indeed, the increase of strain is not producing any change in $pA$, the force by which the solid is responding to the external stimulus. In this region, charge depletion at atomic planes is at it maximum, at the same time  charge accumulation in the interlayer region also changes and is at its global minimum. The analysis of the evolution of the DOS further reveals that first order phase transition is driven by Van-Hove singularities crossing the Fermi level, making available (or unavailable) novel electronic degrees of freedom to the response of the solid to pressure.  

In realistic macroscopic situations many processes take place, such as gliding along crystallographic planes, formation of dislocations, deformation bands, folds etc.., which conceal the microscopic response of the electronic system alone. Our findings unveil the specific way in which the electronic degrees of freedom might contribute to or actually cause the mechanical response of materials. 
The analysis of how the electronic charge within the solid is affected by mechanical stresses makes this study relevant also in the field of tribology. Indeed adhesion and frictional forces are dictated by the electronic charge redistribution occurring when two isolated surfaces are mated to form an interface \cite{Wolloch-18}.

\begin{acknowledgments}
MM would like to thank Prof. Francesco Zamponi and Dr. Paolo Restuccia for fruitful discussions.
These results are part of the "Advancing Solid Interface and Lubricants by First Principles Material Design (SLIDE)" project that has received funding from the European Research Council (ERC) under the European Union's Horizon 2020 research and innovation program (Grant agreement No. 865633). The authors acknowledge funding by the European Union - NextGenerationEU (National Sustainable Mobility Center CN00000023, Italian Ministry of University and Research Decree n. 1033 - 17/06/2022, Spoke 11 - Innovative Materials $\&$ Lightweighting) and by the European Union - NextGenerationEU (Spoke 6 - Mulstiscale Modelling $\&$ Engineering Applications). The opinions expressed are those of the authors only and should not be considered as representative of the European Union or the European Commission’s official position. Neither the European Union nor the European Commission can be held responsible for them. Computational resources were provided by CINECA under the ISCRA initiative and by the CRESCO/ENEAGRID High Performance Computing infrastructure \cite{cresco}.
\end{acknowledgments}


\bibliography{biblio}

\end{document}


\title{SUPPLEMENTAL MATERIAL: On the role played by electrons in the stress-strain curves of ideal crystalline solids}
\maketitle

\section{Structural modifications upon compression}

\begin{figure}[htb!]
\includegraphics[width=0.25\textwidth]{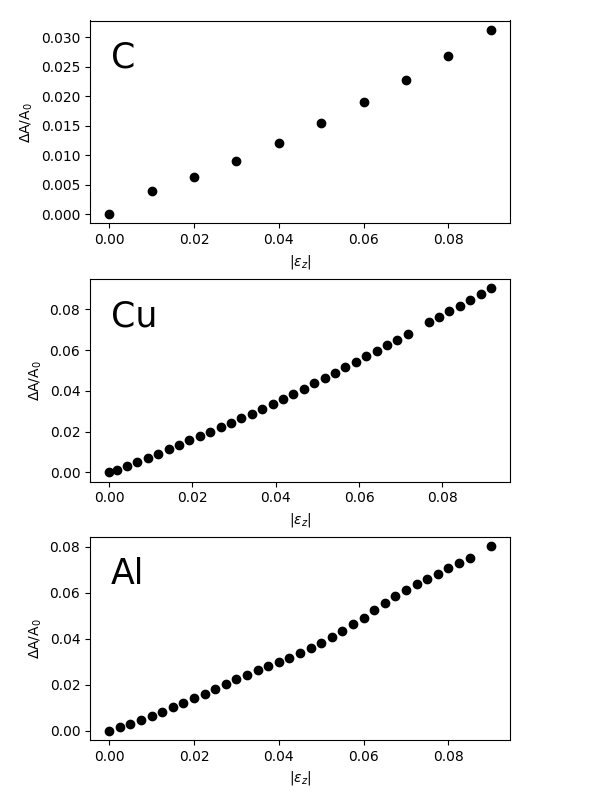}%
\includegraphics[width=0.25\textwidth]{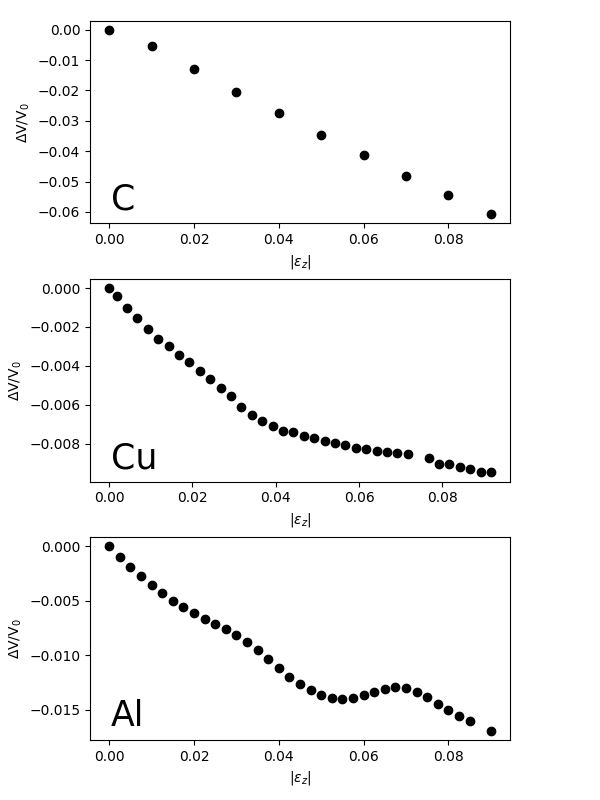}%
\caption{\label{fig:area_vs_strain} Left: variations of the basis cell area $\Delta A/A_0 $ with respect to the vertical strain $|\epsilon_z|$. Right: variations of the cell volume  $\Delta V/V_0 $ with respect to vertical strain $|\epsilon_z|$. $A_0$ and $V_0$ denote the equilibrium area and volume.}
\end{figure}

As shown in Fig. \ref{fig:area_vs_strain}, in all the investigated systems, when the strain $|\epsilon_z|$ is increased, the basis area increases while the cell volumes decreases, consistently with positive isothermal compressibilities 
and in agreement with what has been found for other materials \cite{hou_intermetallics2014,nestola_scirep_2019}. 

As shown in the right panel of Fig. \ref{fig:area_vs_strain}, the change in cell volume is strongly influenced by the material under consideration. In fact, the maximum applied strain $\epsilon_z=$ 0.09 relates to a small volume reduction for aluminum and copper (1.7\%-0.9\%), whereas a larger reduction of 6.3\% is observed for diamond, exhibiting an almost perfectly linear decrease. This can be linked to the stiffness of the diamond lattice which is not able to accomodate strain by deforming and expanding its basis cell area as much as Al and Cu do.  As will be shown in the next section, the higher $|\Delta V/V_0|$ of diamond are associated to much larger stress $|\sigma_{zz}|$ by which this material opposes the given target strains. Importantly, despite the fact that full relaxation is allowed concerning the atomic positions, no significant displacement is found at any strain even for the metals that display the elastic-plastic transition: Cu atoms deviate at most of $0.34\%$ with respect to their ideal position in the $x$ and $y$ directions and 
$0.1\%$ in the $z$ direction; Al atoms maximum relative displacements are $0.16\%$ in the $x$ and $y$ directions and $0.28 \%$ along $z$; whereas the maximum relative displacement of C atoms are of the order of $10^{-5}$ and $10^{-3}$ in the $x$ and $y$, and $z$ direction respectively.

\section{Elastic properties}
The Young’s modulus $E$ has been calculated by fitting the linear region close to the origin of the stress strain curves where the materials obey Hooke’s law $\sigma=E\cdot \epsilon$.  The results are reported in Tab.\ref{tab:elastic_properties} and agree with experimental observations. \par
\begin{figure}[htb!]
\includegraphics[width=0.3\textwidth]{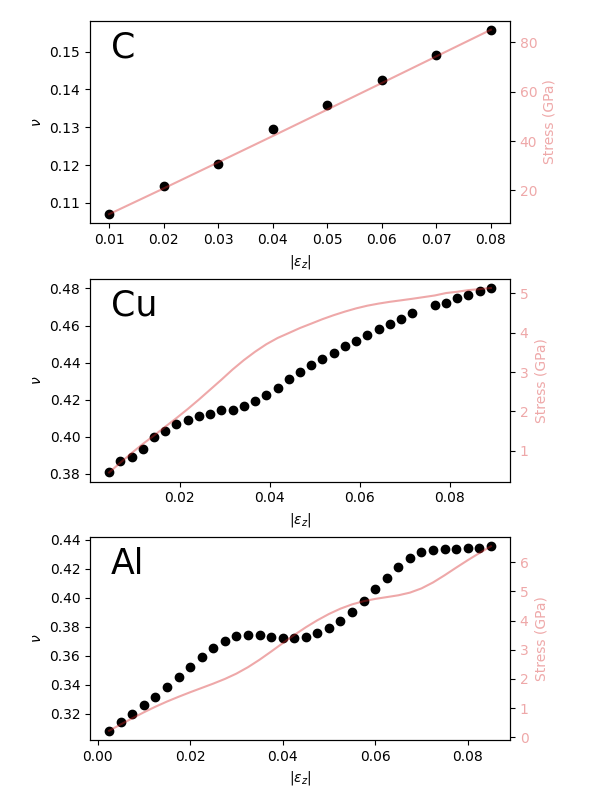}
\caption{\label{fig:poisson-strain} Poisson’s ratio $\nu$ against vertical strain $|\epsilon_z|$ for C (top), Cu (center) and Al(bottom). In red, as a guide to the eye, also the stress values are reported.}
\end{figure}

Poisson’s ratio $\nu$, the ratio between 
transverse and longitudinal (with respect to the load direction) strain, in our case $\nu=-\epsilon_{x}/\epsilon_{z}$, is a measure of how much a material resists in deforming upon uniaxial stress: the higher $\nu$, the higher the material resistance to changes in volume. For a perfectly incompressible, isotropic material the Poisson ratio should reach the ideal value of 0.5. 
$\nu$ provides insights into the nature of atomic bonding and chemical properties of
a solid: Poisson’s ratio values for covalent materials are typically around $\sim$0.1,
$\sim$0.25 for ionic materials, and vary between 0.28 and 0.42 for metals \cite{greaves_natmat2011}. Consistently, the calculated Poisson’s ratio for diamond, in the limit of $0$ strain is $\sim$0.1, reflecting its covalent bonding nature, in agreement with experiments \cite{greaves_natmat2011}, and 
Poisson’s ratios of aluminum and copper are $\sim$0.32 and $\sim$0.34. The computed values are in excellent agreement with experiments as shown in Tab.\ref{tab:elastic_properties}.  
As shown in Fig.\ref{fig:poisson-strain}, Poisson’s ratio increases with pressure for all materials under investigation. Similar trends have been found employing classical force fields for wurtzite aluminum nitrate and diamond  \cite{guler_chinese_j_physics_2014,guler_chinese_j_physics_2015}, and using DFT for Ni$_3$X compounds \cite{hou_intermetallics2014} under load.
\begin{table}[ht!]
    \centering
    \begin{tabular}{c|cc|cc}
    \hline \hline
         Material &  $E$ (GPa) &  $E_{exp}$ (GPa) &  $\nu$ &  $\nu_{exp}$\\
         \hline
         C  & 1068  & 700-1200 & 0.107  & 0.10-0.30\\
         Cu & 97    & 110 & 0.341  & 0.34\\
         Al & 82    & 69 & 0.326  & 0.33\\
         \hline \hline
    \end{tabular}
    \caption{Comparison between calculated and experimental values of Young’s modulus $E$ and Poisson’s ratio $\nu$ for Al, Cu, C. Experimental values are obtained from Ref.\cite{callister_book}.}
    \label{tab:elastic_properties}
\end{table}

\section{Identification of Al critical points}
\begin{figure}[ht!]
\centering
\includegraphics[width=0.4\textwidth]{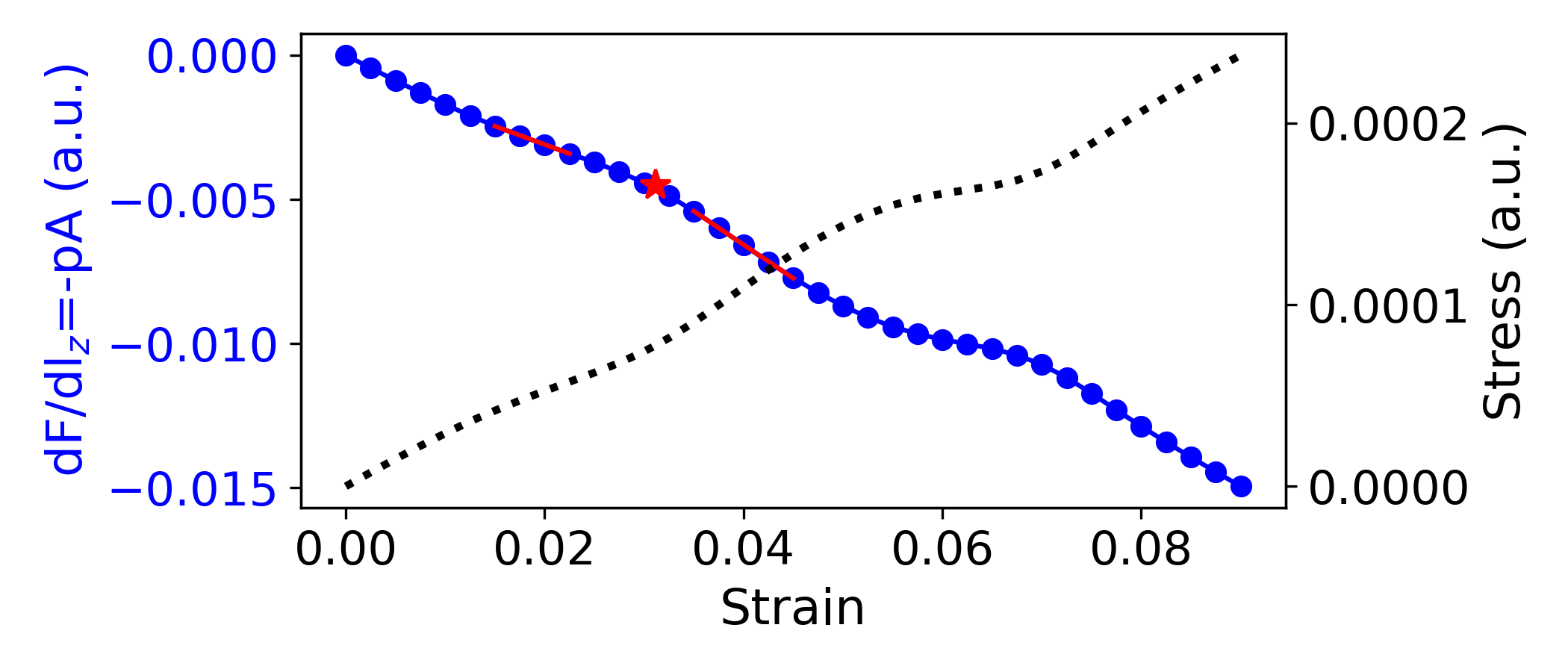}
\includegraphics[width=0.4\textwidth]{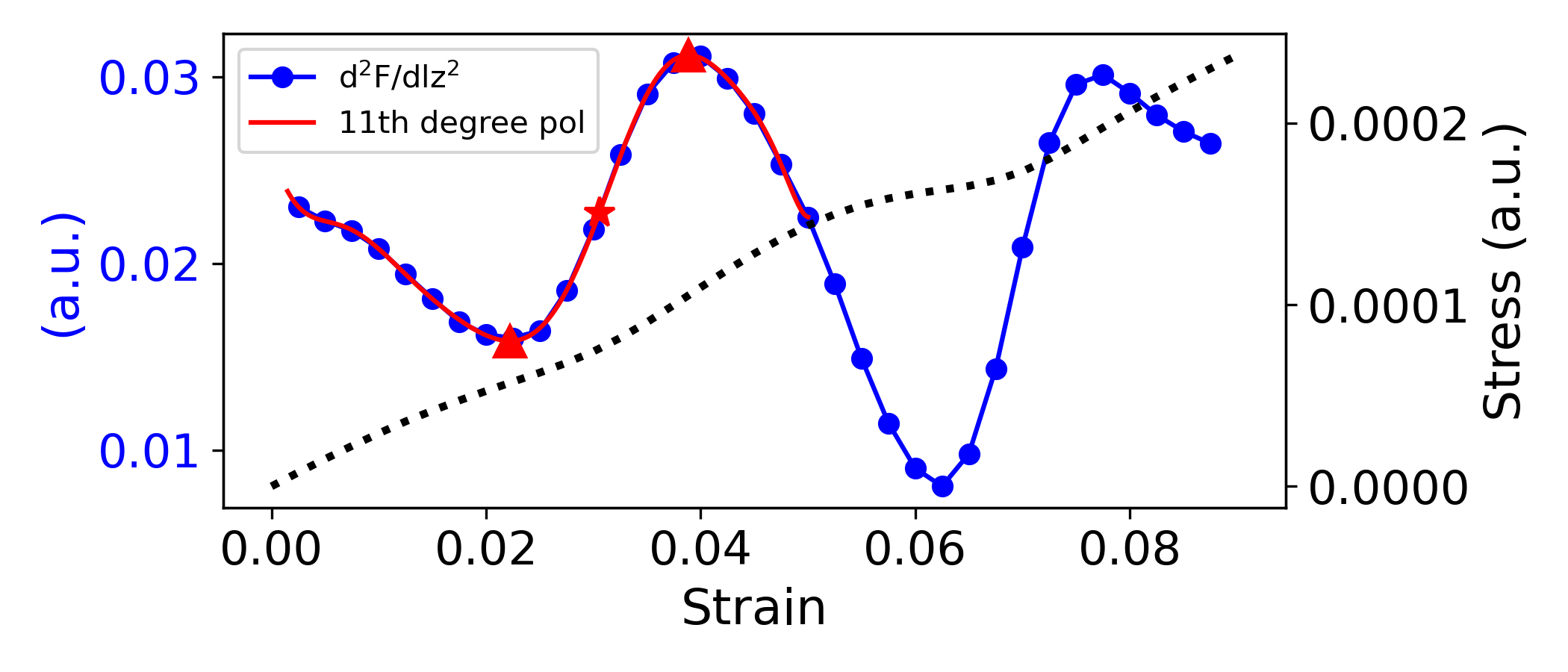}
\caption{\label{fig:second_order} (a) First order derivative of Helmholtz free energy $F$ as a function of strain. Linear fits are shown as red lines. (b) Second order derivative of Helmholtz free energy $F$ as a function of strain. The fit performed with a 11th-degree polinomial is shown as a red line and the local minimum and maximum are marked with red triangles. In all panels the critical point is marked with a red star and stress values are reported as a black dotted line referring to the right axis labels.}
\end{figure}

\begin{figure}[ht!]
{\includegraphics[width=0.4\textwidth]{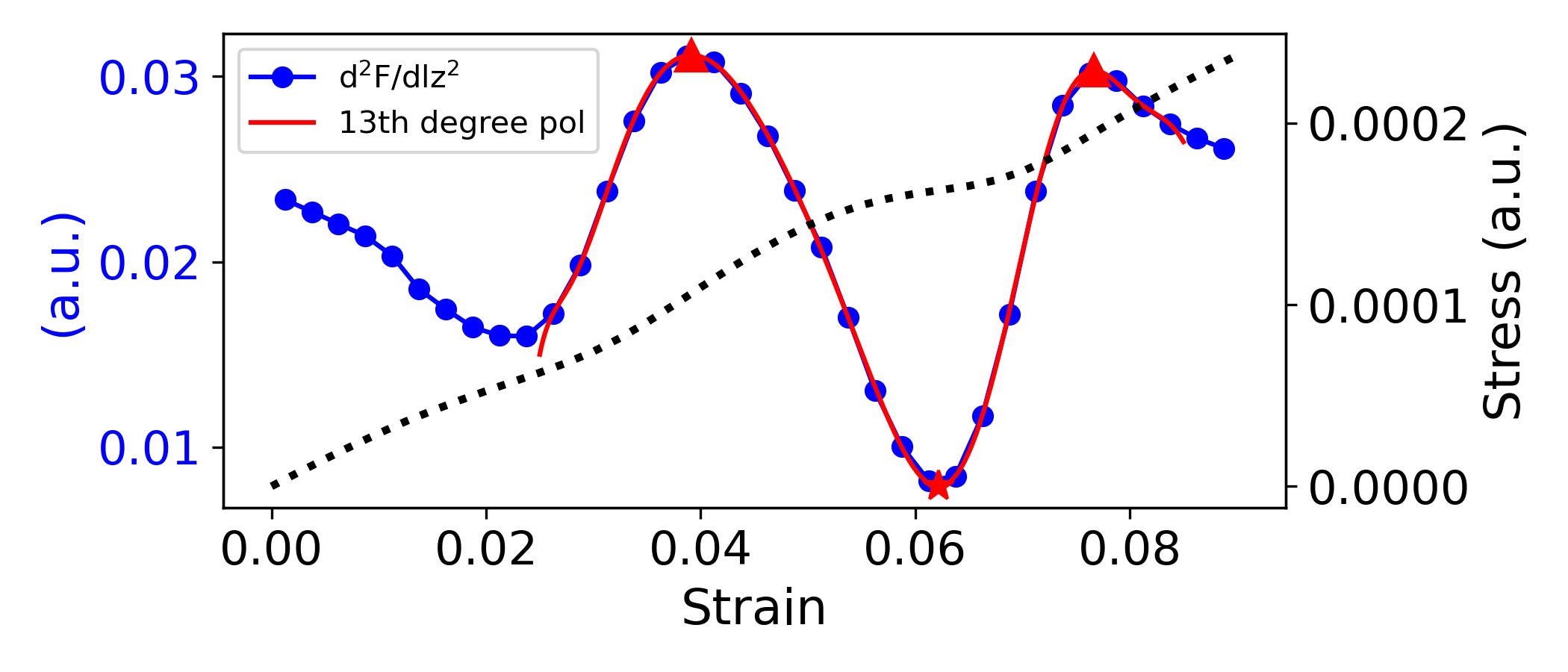}}
\caption{\label{fig:first_order} Second order derivative of Helmholtz free energy $F$ as a function of strain. The fit performed with a 13th-degree polinomial is shown as a red line, the local maxima are marked with red triangles, while the critical point is indicated with a red star. Stress values are reported as a black dotted line referring to the right axis labels.}
\end{figure}

To identify the critical point associated to the second order phase transition of Al, the first and second order derivatives of the thermodynamic potential $F$ are analyzed. From the first derivative of $F$ (Fig.\ref{fig:second_order}(a)), linear regressions are performed over two distinct strain intervals, 0.015$\le$$\epsilon$$\le$0.0225 and 0.035$\le$$\epsilon$$\le$0.045. The critical point is defined as the intersection between the two resulting lines, yielding a value of $\epsilon$=0.03. By fitting the second derivative of $F$ (Fig.\ref{fig:second_order}(b)) with a 11th-degree polynomial in the range 0.0015$\le$$\epsilon$$\le$0.05, the local minimum and maximum of the interpolated curve, occurring at $\epsilon_{min}=0.022$ and $\epsilon_{max}=0.0389$ respectively, are identified. The average of these two values, $\epsilon$=0.03, provides an alternative estimate of the critical point. Both approaches converge to a consistent estimate of the critical point at 0.03 strain, marking the second order phase transition.  \\
The critical point associated to the first order phase transition of Al is identified by interpolating the second derivative of $F$  with a 13th-degree polynomial in the range 0.025$\le$$\epsilon$$\le$0.085 (Fig.\ref{fig:first_order}). The two local maxima at $\epsilon$=0.04 and 0.0775 identify the range of the transition, while the minimum at $\epsilon$=0.0625 represents the critical point.

\section{Al DOS analysis}
\begin{figure}[H]
    \centering
    \includegraphics[width=0.4\textwidth]{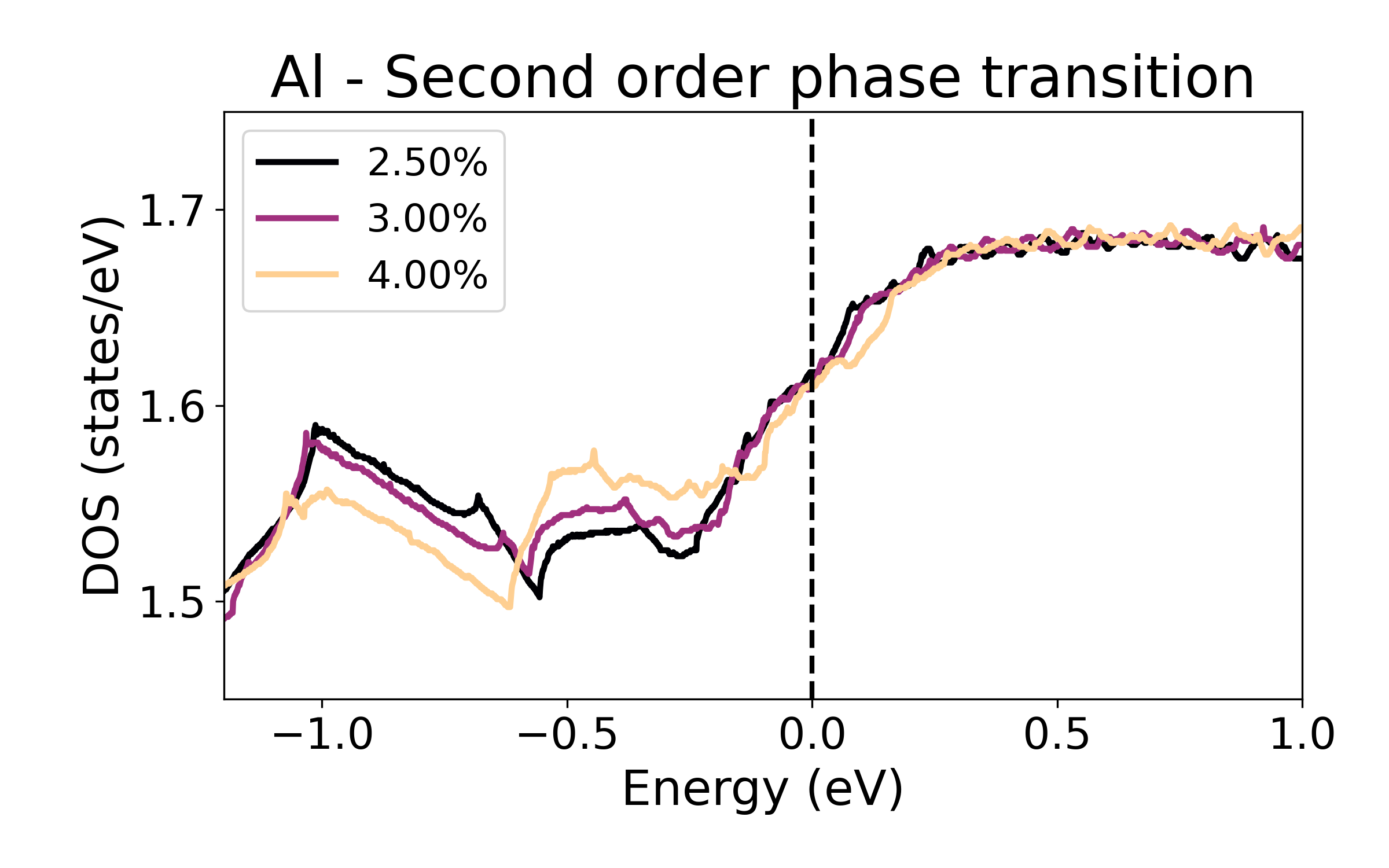}
    \caption{Al DOS for strain values around the critical $\epsilon_z=0.03$.}
    \label{fig:dos_al_0.03_strain}
\end{figure}

\begin{figure}[H]
    \centering
    \includegraphics[width=0.4\textwidth]{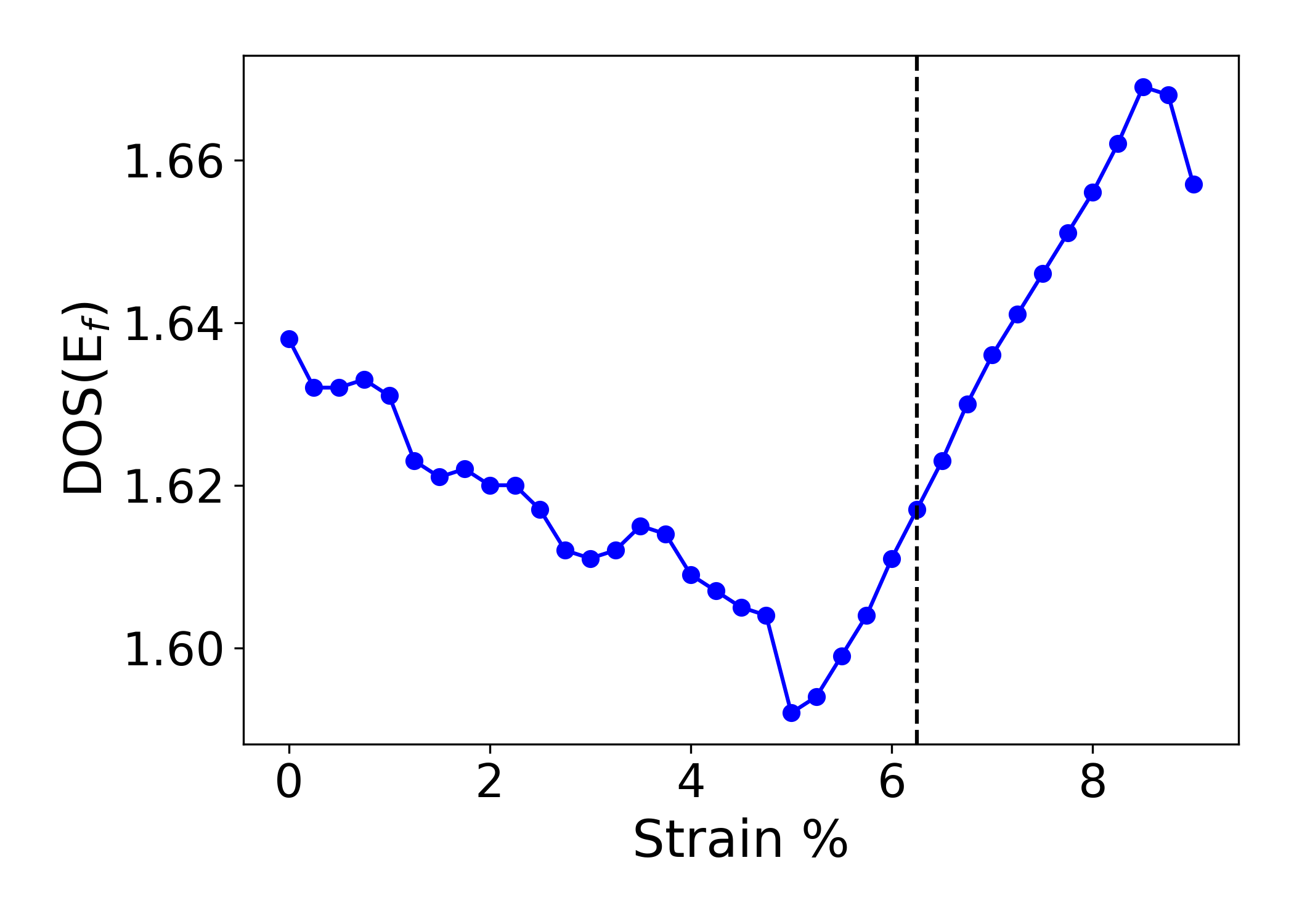}
    \caption{Value of the density of states at the Fermi level of Al as a function of strain.}
    \label{fig:dos_at_fermi}
\end{figure}

The behaviour of Al density of states calculated for different strains around the critical value of 3$\%$, which marks the second order phase transition, is shown in Fig.\ref{fig:dos_al_0.03_strain}. In this case, there are no Van Hove singularities crossing the Fermi level, suggesting that no Lifshitz transition is occurring. \\
In Fig.\ref{fig:dos_at_fermi}, the density of states at the Fermi level of Al is plotted as a function of strain. At $\epsilon = 0.05$, corresponding to the onset of a Van Hove singularity crossing the Fermi level (see Fig.8(b) in the main paper), a discontinuity in $\mathrm{DOS}(E_\mathrm{F})$ can be observed, marking a Lifshitz transition. From this point up to $\epsilon = 0.08$, i.e., throughout the entire first-order phase transition, $\mathrm{DOS}(E_\mathrm{F})$ increases linearly.


\bibliography{biblio}